\documentclass[aps,preprint,nofootinbib,superscriptaddress]{revtex4}

\usepackage{amssymb}

\usepackage{epsfig}
\usepackage[bookmarksnumbered,bookmarksopen,colorlinks,citecolor=blue,linkcolor=blue]{hyperref}

\begin{document}

\title{ Fusion and quasi-fission dynamics in nearly-symmetric reactions }

\author{Ning Wang}
\email{wangning@gxnu.edu.cn}\affiliation{ Department of Physics,
Guangxi Normal University, Guilin 541004, People's Republic of
China }
\affiliation{ State Key Laboratory of Theoretical Physics, Institute of Theoretical Physics, Chinese Academy of Sciences, Beijing 100190, People's Republic of China}

\author{Kai Zhao}
\affiliation{China Institute of Atomic Energy, Beijing 102413, People's Republic of
China}

\author{Zhuxia Li}
\affiliation{China Institute of Atomic Energy, Beijing 102413, People's Republic of
China}

\begin{abstract}

Some nearly-symmetric fusion reactions are systematically investigated with the improved quantum molecular dynamics
(ImQMD) model.  By introducing two-body inelastic scattering in the Fermi constraint procedure, the stability of an individual nucleus and the description of fusion cross sections at energies near the Coulomb barrier can be further improved. Simultaneously, the quasi-fission process in $^{154}$Sm+$^{160}$Gd is also investigated with the microscopic dynamics model for the first time. We find that at energies above the Bass barrier, the fusion probability is smaller than $10^{-5}$ for this reaction, and the nuclear contact-time is generally smaller than $1500$ fm/c. From the central collisions of Sm+Gd, the neutron-rich fragments such as $^{164,165}$Gd, $^{192}$W can be produced in the ImQMD simulations, which implies that the quasi-fission reaction could be an alternative way to synthesize new neutron-rich heavy nuclei.

\end{abstract}
\maketitle

\begin{center}
\textbf{I. INTRODUCTION}
\end{center}

The heavy-ion fusion reaction is an important way not only for the study of the nuclear structures, but also for the synthesis of new heavy and super-heavy nuclei (SHN) \cite{OPb2,OSm,Timm,Zhang10,Hof00,Ogan10,Sob,Gup05}. For relatively light fusion systems, the fusion (capture) cross sections of heavy-ion reactions can be accurately predicted from the fusion coupled channel or barrier distribution calculations based on the barrier penetration concept together with some suitable nucleus-nucleus potentials \cite{Wong73,Hag99,liumin06,Wang09,Wangbin}. To understand the dynamical process in fusion reactions, some microscopical dynamics models, such as the time-dependent Hartree-Fock (TDHF) model \cite{Umar06,Umar12,Guo07} and the improved quantum molecular dynamic (ImQMD) model \cite{ImQMD2002,ImQMD2004,ImQMD2010} have been developed. As an improved version of the quantum molecular dynamics (QMD) model which was proposed for simulating heavy-ion collisions (HICs) at intermediate and high energies, the ImQMD model is successfully applied on heavy-ion fusion reactions between stable nuclei and the reactions induced by neutron-rich nuclei, with a series of modifications aiming at the study of heavy-ion reactions at intermediate and low energies  \cite{ImQMD2012,ImQMD2014,Wang14a}. In the previous work, the ImQMD model is tested for the description of a number of fusion reactions induced by $^{16}$O \cite{Wang14a}. The fusion cross sections for these asymmetric reactions can be well reproduced with the ImQMD model at energies near and above the Coulomb barrier. We also note that for the fusion reactions with heavy target nuclei such as $^{186}$W, the fusion cross sections at sub-barrier energies are over-predicted. It could be due to the influence of few spurious nucleon emission after a relatively longer evolution for the projectile and target nuclei at relatively low incident energies. It is therefore necessary to further improve the model for a better description of the fusion reactions at sub-barrier energies.

In addition to the fusion process, the quasi-fission process in the reactions leading to the synthesis of super-heavy nuclei have also attracted a lot of attentions in recent decades \cite{Shen87,Swia81,Kozu14,Ober14,Zhang10a}. The dynamics models such as the diffusion model based on the master equation \cite{Adam97,Diaz01,Nan12} and the macroscopic fluctuation-dissipation model based on the Langevin equation \cite{Shen02,Zagr05} have been applied for the description of quasi-fission process. Although some measured evaporation residual cross sections of super-heavy nuclei can be reasonably well reproduced, the uncertainty of the predicted fusion probability from these different models for unmeasured systems is still large due to the uncertainty of model parameters such as the fission barrier height of super-heavy nuclei \cite{Nas11,Bao15}. For example, with the fusion-by-diffusion model, the predicted evaporation residual cross sections of $^{154}$Sm+$^{154}$Sm by Choudhury and Gupta \cite{Chou14} are larger than those from Cap et al. \cite{Cap14} by 13 orders of magnitude, by adopting different values for the injection point distance and fission barrier. It is therefore necessary to investigate the quasi-fission dynamics in such kind of reactions with a self-consistent microscopic dynamics model.

\begin{figure}
\includegraphics[angle=0,width=0.6 \textwidth]{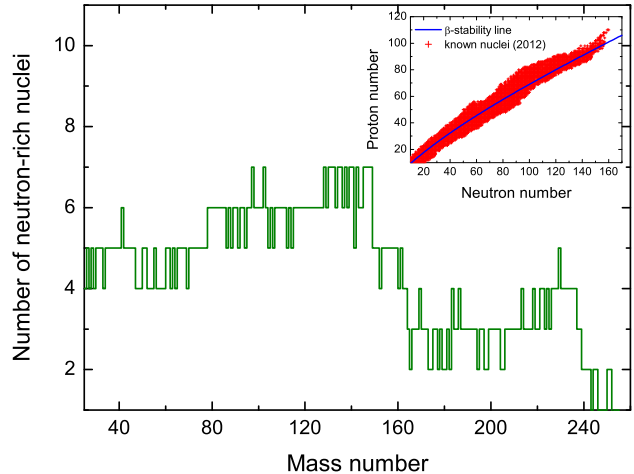}
\caption{(Color online) Number of known neutron-rich nuclei as a function of mass number.   }
\end{figure}

On the other hand, the masses of neutron-rich heavy nuclide are of great importance for the study of nuclear structures and nuclear astrophysics. In Fig. 1, we show the number of neutron-rich nuclei (i.e., those below the $\beta$-stability line) with known masses as a function of mass number $A$. The crosses in the sub-figure give the positions of known nuclei in atomic mass evaluation AME2012 \cite{Audi12}, and the solid curve denotes the $\beta$-stability line described by Green's formula. One sees that at neutron-rich side, the number of known nuclei decreases obviously at mass region $A>160$. This is because the synthesis of new neutron-rich heavy nuclei through heavy-ion fusion reactions is limited due to the neutron numbers of available stable nuclei as the projectile and target, similar to the difficulties in the synthesis of neutron-rich SHN. One requires an alternative way to synthesize the new neutron-rich nuclei (especially the ones with mass number $ 165\le A \le 205 $), in order to measure their masses. It is known that large charge and mass transfer occurs in the quasi-fission process, which might provide a chance to produce new neutron-rich heavy nuclide. It is therefore interesting to study the new nuclide in the quasi-fission fragments.

The structure of this paper is as follows: In sec. II, the framework of the ImQMD model, especially the modification of the Fermi constraint will be  introduced. In sec. III, the fusion cross sections of six nearly-symmetric reactions are systematically calculated to test the model reliability and the closest distance between two nuclei in back-angle scattering is also studied. In sec. IV, the quasi-fission dynamics and fusion probability of $^{154}$Sm+$^{160}$Gd will be further investigated with the improved version of the ImQMD model. Finally a brief summary is given in Sec. V.

\begin{center}
\noindent{\bf {II. IMPROVED QUANTUM MOLECULAR DYNAMICS MODEL }}\\
\end{center}

In the improved quantum molecular dynamics model, each
nucleon is represented by a coherent state of a Gaussian wave
packet. The density distribution function $\rho$ of a system reads
\begin{equation} \label{1}
\rho(\mathbf{r})=\sum_i{\frac{1}{(2\pi \sigma_r^2)^{3/2}}\exp
\left [-\frac{(\mathbf{r}-\mathbf{r}_i)^2}{2\sigma_r^2} \right ]},
\end{equation}
where $\sigma_r$ represents the spatial spread of the wave packet.
The propagation of nucleons is governed by the self-consistently generated mean field,
\begin{equation} \label{2}
\mathbf{\dot{r}}_i=\frac{\partial H}{\partial \mathbf{p}_i}, \; \;
\mathbf{\dot{p}}_i=-\frac{\partial H}{\partial \mathbf{r}_i},
\end{equation}
where $r_i$ and $p_i$ are the center of the $i$-th wave packet in
the coordinate and momentum space, respectively.  The Hamiltonian
$H$ consists of the kinetic energy
$T=\sum\limits_{i}\frac{\mathbf{p}_{i}^{2}}{2m}$ and the effective
interaction potential energy $U$:
\begin{equation} \label{3}
H=T+U.
\end{equation}
The effective interaction potential energy is written as the sum of
the nuclear interaction potential energy $U_{\rm
loc}=\int{V_{\rm loc}(\textbf{r})d\textbf{r}}$ and of the Coulomb
interaction potential energy $U_{\rm Coul}$ which includes the
contribution of the direct and exchange terms,
\begin{equation}
U=U_{\rm loc}+U_{\rm Coul}.
\end{equation}
Where $V_{\rm loc}(r)$ is the potential energy density that is
obtained from the effective Skyrme interaction, in which the spin-orbit term is not involved:
\begin{equation}
V_{\rm
loc}=\frac{\alpha}{2}\frac{\rho^2}{\rho_0}+\frac{\beta}{\gamma+1}\frac{\rho^{\gamma+1}}{\rho_0^{\gamma}}+\frac{g_{\rm
sur}}{2\rho_0}(\nabla\rho)^2
+g_{\tau}\frac{\rho^{\eta+1}}{\rho_0^{\eta}}+\frac{C_s}{2\rho_0}[\rho^2-k_s(\nabla\rho)^2]\delta^2
\end{equation}
where $\delta=(\rho_n -\rho_p)/(\rho_n +\rho_p)$ is the isospin
asymmetry. In Table I we list the model parameters IQ3a  adopted in the calculations. The corresponding value of the incompressibility coefficient of nuclear matter is about 225 MeV.

\begin{table}
 \caption{ Parameter set IQ3a \cite{ImQMD2014}.}
\begin{tabular}{lccccccccccc}
\hline Parameter & $\alpha $ & $\beta $ & $\gamma $ &$%
g_{\rm sur}$ & $ g_{\tau }$ & $\eta $ & $C_{s}$ & $\kappa _{s}$ &
$\rho
_{0}$ & ~~$\sigma_0$~~ & ~~$\sigma_1$~~ \\
 & (MeV) & (MeV) &  & (MeVfm$^{2}$) & (MeV) &  & (MeV) & (fm$^{2}$) &
 (fm$^{-3}$) & (fm) & (fm) \\ \hline
IQ3a & $-207$ & 138 & 7/6 & 16.5 & 14 & 5/3 &  34  & 0.4  & 0.165 & 0.94 & 0.02\\
  \hline
\end{tabular}
\end{table}

To describe the fermionic nature of the $N$-body system and to improve the stability of an individual nucleus, the Fermi constraint is simultaneously adopted in the ImQMD model. In the version ImQMD-v2.1 \cite{ImQMD2014}, the phase space occupation numbers are checked during the propagation of nucleons. If $\bar f_{i}>1$, the momentum of the particle $i$ are randomly changed by a series of two-body elastic scattering between this particle and its neighboring particles, similar to those do in the CoMD model \cite{constrain}. The Pauli blocking condition and the total energy of the system at the next time step are simultaneously checked. The Fermi constraint affects the low momentum part of the momentum distribution strongly, and can effectively improve the momentum distribution at low momentum region. However, the standard Fermi constraint approach does not improve the long tail (high momentum part) of the momentum distribution. With ImQMD-v2.1, the average numbers of spurious emitted nucleons at $t=2000$ fm/c are 1.1 for the individual $^{92}$Zr nuclei and 2.6 for $^{132}$Sn with the parameter set IQ3a, respectively. To further improve the stability of an individual nucleus, we modify the Fermi constraint procedure in version ImQMD-v2.2. In the new version, we simultaneously consider the two-body inelastic scattering in the Fermi constraint in addition to the elastic scattering involved in v2.1. If the difference between the momentum of a nucleon and that of its neighboring nucleons is larger than Fermi momentum, $|\vec{p}_1-\vec{p}_2|>p_F$, with the Fermi momentum $p_F=260$ MeV/c, a tiny part of momentum $\vec{p} f_t $ of the nucleon with a higher momentum will be transferred to the other one. For heavy-ion fusion reactions, we set the transfer factor $f_t=5\times10^{-6}$ which guarantee that the total momentum and energy of the system are well conserved in the simulations. We find that the consideration of the inelastic scattering in the Fermi constraint, which is helpful to improve the distribution of the phase space occupation in nuclei, can significantly reduce the number of spurious emitted nucleons. Without any additional selection for the initial nuclei, the average numbers of spurious emitted nucleons at $t=2000$ fm/c are reduced to 0.56 for $^{92}$Zr and 1.75 for $^{132}$Sn, respectively. The initialization of the ImQMD simulations is as the same as those adopted in Refs. \cite{ImQMD2014,Wang14a} and the collision term is not involved in the present calculations. In addition, the new version ImQMD-v2.2 has also been tested for the description of multi-fragmentation process for heavy-ion collision at intermediate energies. The charge distribution for central collisions of $^{197}$Au+$^{197}$Au at an incident energy of 35 AMeV can be well reproduced with this version.

\begin{center}
\textbf{III. Fusion cross sections and dynamics in intermediate reaction systems}
\end{center}

In this section, the fusion excitation functions of $^{16}$O+$^{186}$W and $^{132}$Sn+$^{48}$Ca are firstly re-examined with the version ImQMD-v2.2. Then,
the fusion reactions $^{28}$Si+$^{30}$Si, $^{32}$S+$^{48}$Ca, $^{40}$Ca+$^{48}$Ca, and $^{86}$Kr+$^{76}$Ge will be systematically investigated to test the model reliability for describing light and intermediate fusion systems.

\begin{figure}
\includegraphics[angle=0,width=1\textwidth]{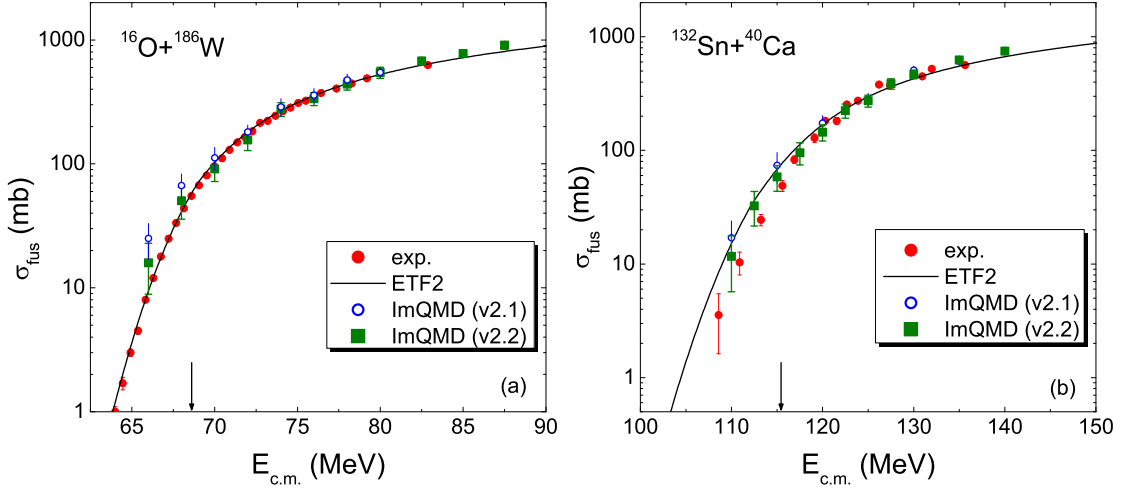}
\caption{(Color online) Fusion excitation functions of $^{16}$O+$^{186}$W and $^{132}$Sn+$^{40}$Ca. The solid circles denote the experimental data taken from Refs. \cite{OSm,CaSn}. The solid curves denote the results with an empirical barrier distribution in which the fusion barrier is obtained by using the Skyrme energy-density functional together with the extended Thomas-Fermi (ETF2) approximation \cite{liumin06,Wang09}. The solid squares and open circles denote the results of ImQMD with version v2.2 and v2.1, respectively. The statistical errors in the ImQMD calculations are given by the error bars. The arrow denotes the position of the most probable barrier height. }
\end{figure}

Through creating certain bombarding events at each incident energy $E_{\rm c.m.}$ and each impact parameter $b$, and counting the number of fusion events, we obtain the fusion probability $g_{\rm fus}(E_{\rm c.m.},b)$ for a certain fusion reaction. The corresponding fusion excitation function can be calculated with
\begin{equation}
\sigma _{\rm fus}(E_{\rm c.m.})=2\pi \int b \, g_{\rm fus} \, db
\simeq 2\pi \sum b \, g_{\rm fus} \, \Delta b.
\end{equation}
Where, we set $\Delta b=1$ fm. In the calculation of the fusion probability for light and intermediate reaction system, the event is counted as a fusion event if the center-to-center distance between the two nuclei is smaller than the nuclear radius of the compound nuclei. Without introducing any free model parameters and/or additional assumptions, the whole reaction process for all reactions mentioned is self-consistently simulated with this semi-classical microscopic dynamics model.

\begin{figure}
\includegraphics[angle=0,width=1\textwidth]{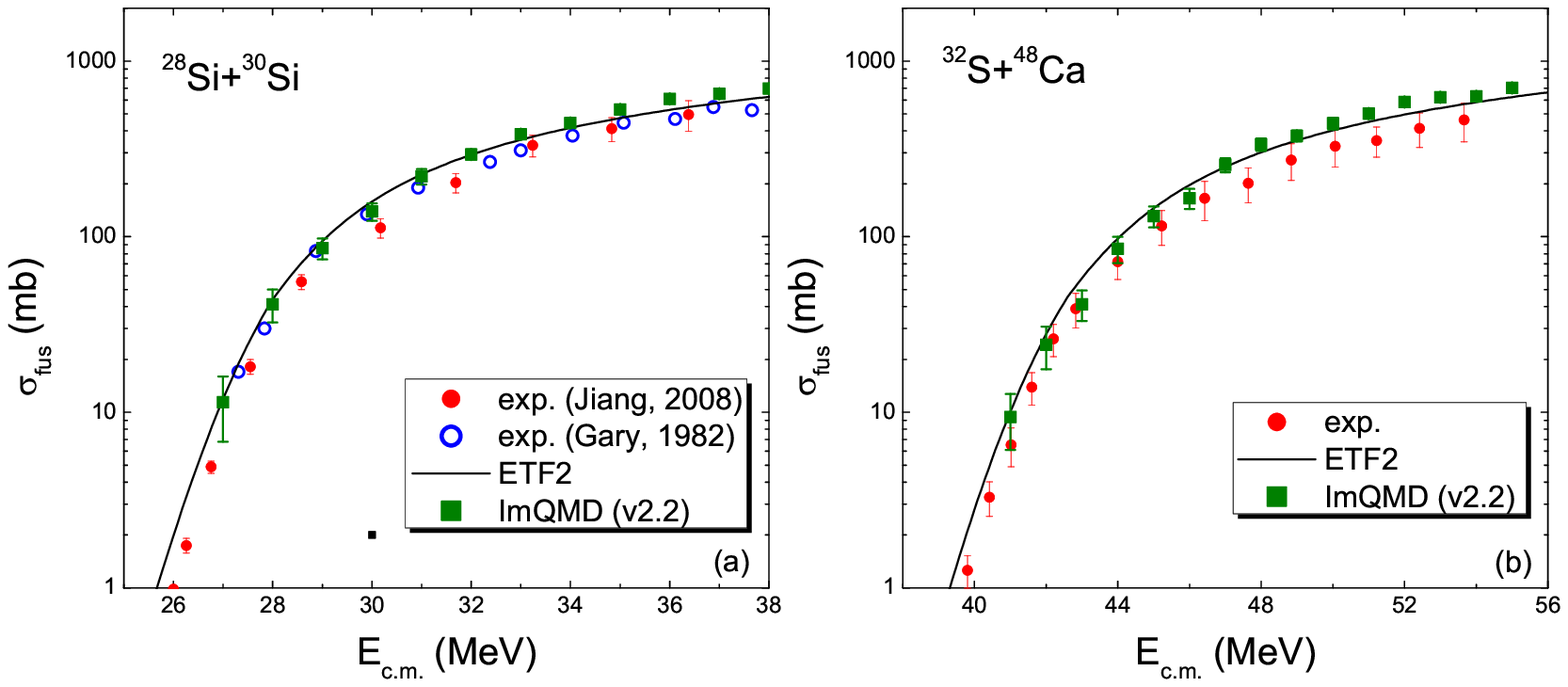}
\caption{(Color online) The same as Fig.1, but for reactions $^{28}$Si+$^{30}$Si \cite{SiSi08,SiSi82} and $^{32}$S+$^{48}$Ca \cite{SCa}.   }
\end{figure}

To test the new version of the ImQMD model for the description of fusion reactions, we re-examine the fusion cross sections of $^{16}$O+$^{186}$W and $^{132}$Sn+$^{40}$Ca. With the previous version ImQMD-v2.1, the fusion cross sections of these two reactions at energies near and above the Coulomb barrier can be well reproduced, whereas the results at sub-barrier fusion are over-predicted \cite{Wang14a}. The spurious emitted nucleons enhance the surface diffuseness of nuclei and thus over-predict the fusion cross sections at sub-barrier energies. With the inelastic-scattering in the Fermi constraint being considered, the average numbers of spurious emitted nucleons in the new version ImQMD-v2.2 are reduced by $50\%$ for $^{92}$Zr and $33\%$ for $^{132}$Sn comparing with those from v2.1, respectively. In Fig. 2, we show the fusion excitation functions of $^{16}$O+$^{186}$W and $^{132}$Sn+$^{40}$Ca. The open circles and solid squares denote the results with the version v2.1 and v2.2, respectively. The solid circles denote the experimental data. One sees that for cross sections at above-barrier energies, the results from v2.1 and v2.2 are almost the same. Whereas, for fusion cross sections at energies around the barrier (with $\sigma_{\rm fus}\simeq 50$ mb) and sub-barrier energies, the results from v2.2 look much better, since the better initial nuclei are obtained in the simulations.

With the same version of the ImQMD model, some nearly-symmetric fusion reactions are also investigated. Fig. 3 shows the calculated fusion excitation functions for  $^{28}$Si+$^{30}$Si and $^{32}$S+$^{48}$Ca. The data in the region ($ \sigma_{\rm fus}>$ 10 mb) can be reasonably well reproduced by ImQMD-v2.2. For deep sub-barrier fusion, it is still difficult for the present version of ImQMD to accurately describe, because the very rare spurious emitted nucleons and microscopic shell and pairing effects might influence the fusion cross sections at low energies.

\begin{figure}
\includegraphics[angle=0,width=0.65 \textwidth]{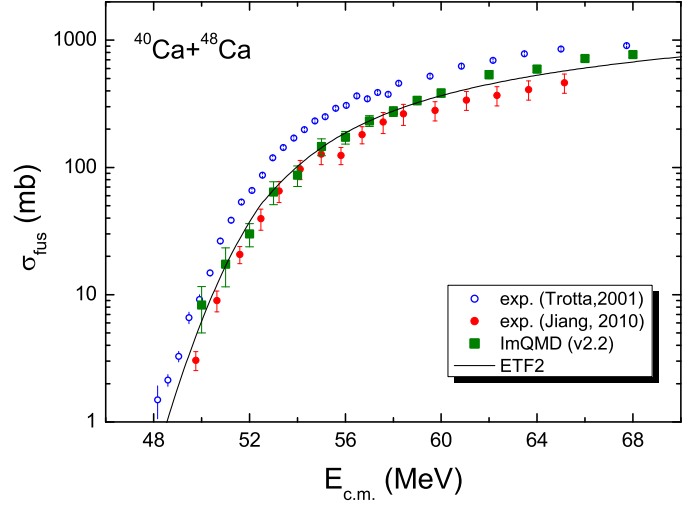}
\caption{(Color online) The same as Fig.1, but for reaction  $^{40}$Ca+$^{48}$Ca. The experimental data are taken from Refs. \cite{CaCa01,CaCa10}. }
\end{figure}

Here, we also study the fusion reactions $^{40}$Ca+$^{48}$Ca and $^{86}$Kr+$^{76}$Ge. For $^{40}$Ca+$^{48}$Ca, the fusion excitation functions have been measured by \emph{Trotta et al}. in 2001 \cite{CaCa01} and re-measured by \emph{Jiang et al}. in 2010 \cite{CaCa10}. It is found that the data from \emph{Trotta et al.} are larger than those from the other group by about a factor of two. It is therefore interesting to study this reaction with some theoretical models. In Fig. 4, we show the predicted fusion cross sections of $^{40}$Ca+$^{48}$Ca at energies around the Coulomb barrier. The solid squares denote the results of ImQMD-v2.2 and the solid curve denote the ETF2 calculations. We find that at energies near the barrier (with $\sigma _{\rm fus}\approx 50-100$ mb) the theoretical predictions from the two different models are close to the data from \emph{Jiang et al}.

\begin{figure}
\includegraphics[angle=0,width=0.65 \textwidth]{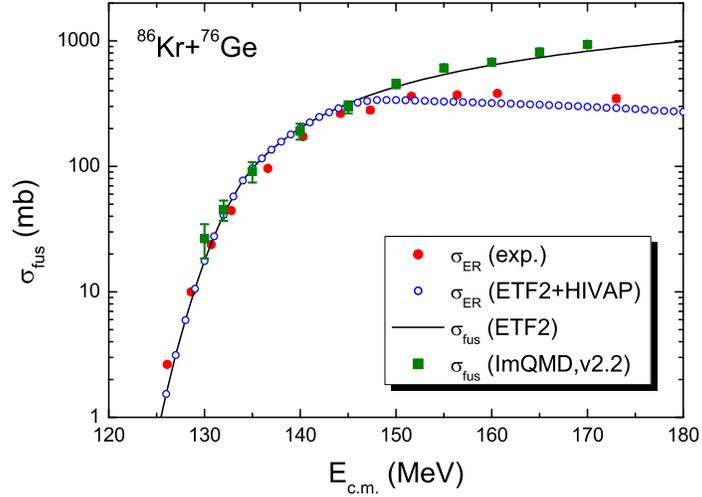}
\caption{(Color online) Fusion excitation functions of $^{86}$Kr+$^{76}$Ge \cite{Reis85}. The solid circles and open circles denote the measured and predicted evaporation residual cross sections, respectively.}
\end{figure}

It is generally thought that for systems with the charge number $Z_{\rm CN}$ of compound nuclei smaller than about 60, the fission barrier is high enough to make fission an improbable decay mode at incident energies close to the fusion barrier \cite{Reis85,Wang09}. Thus for these reactions, the evaporation residual cross section $\sigma_{\rm ER}$ approximately equal to the fusion cross section $\sigma_{\rm fus}$ at near-barrier energies. For heavier compound systems the fission increases rapidly with the $Z_{\rm CN}^{\; 2}/A_{\rm CN}$ and the angular momentum. For fusion-fission systems, it is generally recognized that $\sigma_{\rm fus}=\sigma_{\rm ER}+\sigma_{\rm FF}$. Here, $\sigma_{\rm fus}$, $\sigma_{\rm ER}$ and $\sigma_{\rm FF}$ denote the cross sections for fusion, evaporation residue and fission, respectively. For $^{86}$Kr+$^{76}$Ge, the fission cross sections can not be ignored at energies above the Coulomb barrier. In Fig. 5, we show the evaporation residual cross sections and fusion cross sections of $^{86}$Kr+$^{76}$Ge. The solid curve denotes the calculated fusion cross sections from the ETF2 approach. Together with the statistical model HIVAP code \cite{Reis81,Wang08,Wang11}, the evaporation residual cross sections $\sigma_{\rm ER}$ are predicted and presented in the figure (open circles). The measured $\sigma_{\rm ER}$ can be remarkably well reproduced by the model predictions. It is unpractical to self-consistently describe the whole fission process of the compound nucleus by using the microscopic dynamics model, because of the very large time scale for the fusion-fission process. The fusion excitation function of $^{86}$Kr+$^{76}$Ge is also calculated with the ImQMD model, and the results are presented in Fig. 5 for comparison. We find that the results from the microscopic dynamics model and those from the static model ETF2 are close to each other.

\begin{figure}
\includegraphics[angle=0,width=0.65 \textwidth]{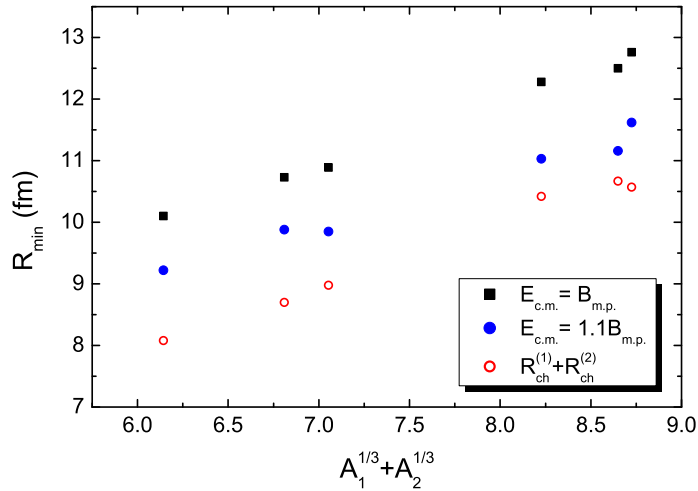}
\caption{(Color online) Closest distance between two nuclei from back-angle quasi-elastic scattering simulations. $R_{\rm ch}^{(1)}$ and $R_{\rm ch}^{(2)}$ denote the measured charge radii \cite{Ang13} of projectile and targets, respectively.}
\end{figure}

To understand the dynamical effects, we also investigate the smallest distance $R_{\rm min}$ between two nuclei in the back-angle quasi-elastic scattering events. Fig. 6 shows the predicted smallest distance $R_{\rm min}$ in the six fusion reactions studied previously as a function of $A_1^{1/3}+A_2^{1/3}$, where $A_1$ and $A_2$ are the mass number of projectile and target nuclei, respectively. The open circles denote the sum of the measured charge radii for the corresponding projectile and target nuclei. The solid squares and circles denote the obtained $R_{\rm min}$ at incident energies $E_{\rm c.m.}=B_{\rm m.p.}$ and $E_{\rm c.m.}=1.1 B_{\rm m.p.}$, respectively. Here, $B_{\rm m.p.}$ is the most probable barrier height of the fusion system. At $E_{\rm c.m.}=1.1 B_{\rm m.p.}$, the probability of quasi-elastic scattering is about $5\%$ for head-on collisions. One sees that the obtained $R_{\rm min}$ increase linearly with the sum of nuclear sizes in general. The obtained $R_{\rm min}$ is systematically larger than the corresponding value of $R_{\rm ch}^{(1)}+R_{\rm ch}^{(2)}$ by about 2 fm at $E_{\rm c.m.}=B_{\rm m.p.}$, and by about 1 fm at $E_{\rm c.m.}=1.1 B_{\rm m.p.}$, respectively. The decrease of $R_{\rm min}$ with incident energy in the ImQMD simulations is consistent with the energy dependence of the injection point distance, $s_{inj}$, adopted in the fusion-by-diffusion model \cite{Cap14}. The dynamical effects in the heavy-ion fusion reactions cause not only the energy-dependence of $R_{\rm min}$, but also the energy-dependence of the barrier height and positions \cite{ImQMD2014}.

\begin{center}
\textbf{IV. Quasi-fission dynamics in $^{154}$Sm+$^{160}$Gd}
\end{center}

For fusion reactions leading to the synthesis of super-heavy nuclei (SHN), the quasi-fission process significantly influences the formation of SHN. Quasi-fission is characterized by nuclear contact-times that are usually greater than 5 zs (i.e. 1500 fm/c) but much shorter than typical fusion-fission times which require the formation of a compound nucleus \cite{Shen87,Ober14}. Furthermore, in the quasi-fission process much more nucleons are transferred in the contact-time comparing with quasi-elastic scattering process. To illustrate the quasi-fission more clearly, we show in Fig. 7 the time evolution of root-mean-square (rms) matter radii of the reaction system $^{40}$Ca+$^{238}$U for three typical events in the ImQMD simulations. At the same incident energy $E_{\rm c.m.}=200$ MeV which is close to the Coulomb barrier and the same impact parameter $b=1$ fm, the time evolutions of the rms radii of the three typical events are quite different. For the event marked by the open circles, the $R_{\rm rms}$ reaches the smallest value at about $t=500$ fm/c,  and then the dinuclear system (DNS) quickly splits up and the value of $R_{\rm rms}$ increases rapidly with time evolution. For the second event (marked by the solid circles), one sees that the dinuclear system remains a contact-configuration about 2500 fm/c and then split into two fragments. For the third event (with solid curve), the rms radii of the reaction system keep a constant value $R_{\rm rms}\approx A_{\rm CN}^{1/3}$ in general after 1000 fm/c, which indicates that the composite system remains nearly-spherical shapes at least 5000 fm/c. According to the time evolution of rms radii which closely relates to the nuclear contact-times, we can discriminate the fusion events from the quasi-elastic scattering and quasi-fission events in the microscopic dynamics simulations.

 \begin{figure}
\includegraphics[angle=0,width=0.65 \textwidth]{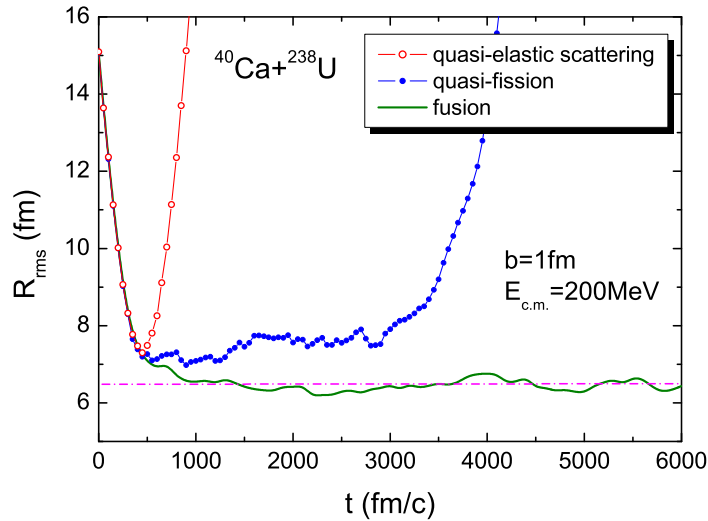}
\caption{(Color online) Time evolution of the rms radii of the system $^{40}$Ca+$^{238}$U for three typical events. The dot-dashed line with $R_{\rm rms}= A_{\rm CN}^{1/3}$ is to guide the eyes.}
\end{figure}

\begin{figure}
\includegraphics[angle=0,width=0.65 \textwidth]{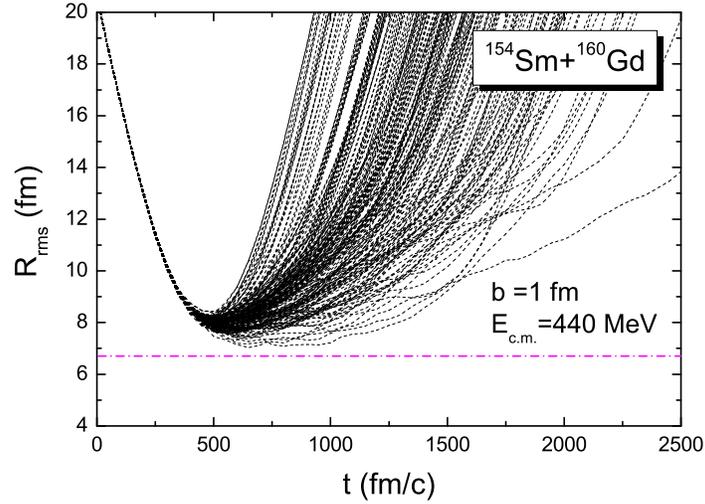}
\caption{(Color online) Time evolution of the rms radii for the reaction system $^{154}$Sm+$^{160}$Gd.}
\end{figure}

\begin{figure}
\includegraphics[angle=0,width=0.7\textwidth]{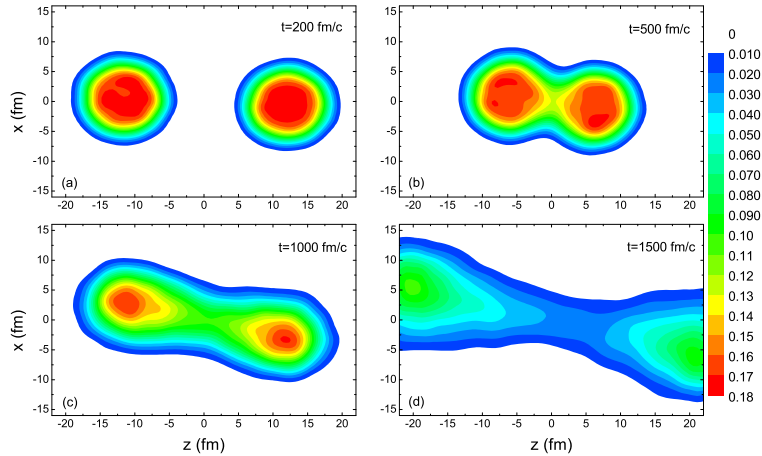}
\caption{(Color online) Time evolution of the density distribution for $^{154}$Sm+$^{160}$Gd at $E_{\rm c.m.}=440$ MeV and $b=1$ fm.}
\end{figure}

For the nearly-symmetric reaction $^{154}$Sm+$^{160}$Gd, the capture pocket in the nucleus-nucleus potential from the ETF2 calculations completely disappears, and therefore the barrier penetration concept is not applicable to describe such a reaction. The Bass barrier \cite{Bass} for $^{154}$Sm+$^{160}$Gd is 393 MeV and the predicted Q-value for complete fusion is $Q=-410$ MeV according to the WS4 calcultions \cite{Wang14}. In this work, we investigate the collisions of $^{154}$Sm+$^{160}$Gd at $E_{\rm c.m.}=440$ MeV which is higher than the Bass barrier by a factor of 1.1. Fig. 8 shows the time evolution of $R_{\rm rms}$ for 200 simulation events. For $^{154}$Sm+$^{160}$Gd, the corresponding value of $A_{\rm CN}^{1/3}=6.8$ fm is given by the dot-dashed horizontal line in the figure. For all 200 events, the values $R_{\rm rms}$ increase after 1000 fm/c even the incident energy is higher than the Bass barrier, and the smallest $R_{\rm rms}$ is about 8 fm which is significantly larger than the rms radii of the compound nuclei ($A=314$, $Z=126$) at spherical shape. From the time evolution of rms radii in Fig. 8, it is difficult to unambiguously discriminate the quasi-fission events from the quasi-elastic scattering ones. To see the dynamical process more clearly, we show in Fig. 9 the time evolution of density distributions. One sees that at $t=500$ fm/c (with smallest $R_{\rm rms}$ in general), the DNS is formed but the density of neck is obviously smaller than the normal density. During $t=500$ to 1000 fm/c most of DNS gradually elongate and tend to split up. After $t=1500$ fm/c, almost all DNS split up. The scattering angles and the breakup time of the DNS are different (see Fig. 8) for different events due to the dynamical effects, which causes the average densities in Fig. 9 at $t=1500$ fm/c is much lower than those at the initial time.

Considering the huge difference of the fusion probability for $^{154}$Sm+$^{160}$Gd from the model predictions, it is necessary to perform a relatively quantitative calculation of the fusion probability based on the ImQMD model. In this work, we create $10^5$ simulation events for $^{154}$Sm+$^{160}$Gd at $E_{\rm c.m.}=440$ MeV and $b=1$ fm. Fig. 10 shows the numbers of DNS as a function of evolution time. At $t=3000$ fm/c, we do not observe any fusion event from all $10^5$ simulation events. In other words, the predicted fusion probability $P_{\rm fus}< 10^{-5}$ for $^{154}$Sm+$^{160}$Gd at $E_{\rm c.m.}=440$   MeV according to the ImQMD-v2.2 calculations. Most of DNS split up after $t=1500$ fm/c. We also note that  the quasi-fission time scale in $^{40}$Ca+$^{238}$U is significantly larger than the corresponding value of $^{154}$Sm+$^{160}$Gd, which is probably due to that the capture pocket of $^{40}$Ca+$^{238}$U is much deeper than that of $^{154}$Sm+$^{160}$Gd according to the ETF2 calculations together with the frozen density approximation \cite{liumin06}.

\begin{figure}
\includegraphics[angle=0,width=0.7 \textwidth]{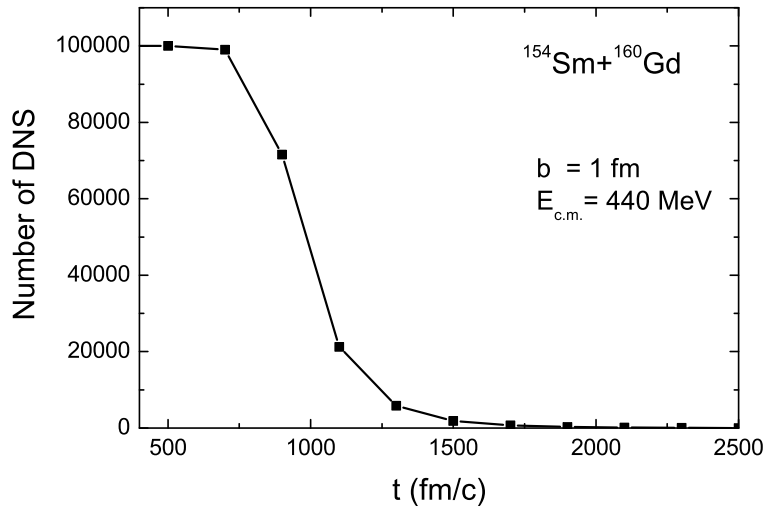}
\caption{(Color online) Number of dinuclear systems with contact-configuration as a function of time.}
\end{figure}

\begin{figure}
\includegraphics[angle=0,width=0.7 \textwidth]{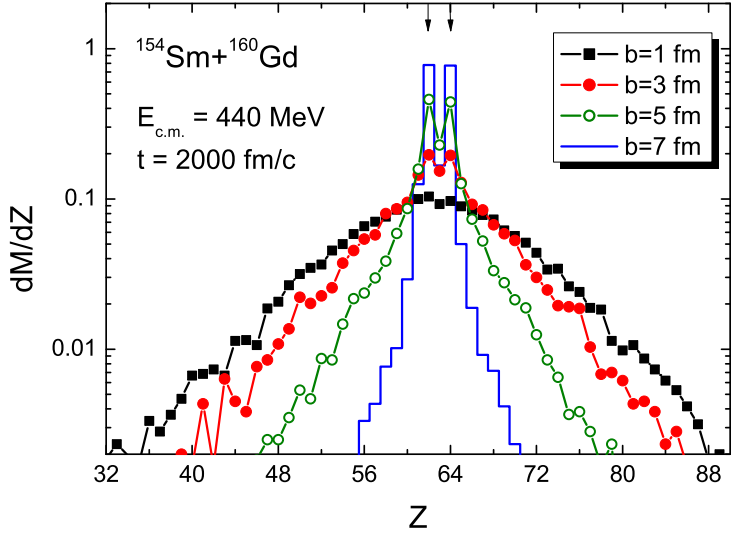}
\caption{(Color online) Charge distribution of fragments in $^{154}$Sm+$^{160}$Gd at $t=2000$ fm/c.}
\end{figure}

Although it is almost impossible to produce super-heavy nuclei in $^{154}$Sm+$^{160}$Gd considering the very small fusion probability, it might produces new neutron-rich nuclide during the quasi-fission process. Here, we study the charge distribution in $^{154}$Sm+$^{160}$Gd from central to peripheral  collisions. Fig. 11 shows the predicted charge distribution at $E_{\rm c.m.}=440$ MeV and at different impact parameters. The arrows denote the charge numbers of the projectile and target nuclei. One sees that for peripheral collisions $b=7$ fm, the charge distribution is narrow with peaks located at the positions of the arrows, which implies that the quasi-elastic scattering plays a dominant role at $b=7$ fm. With the decrease of impact parameter, the width of the charge distribution evidently increase. For the central collisions, the large charge and mass transfer in the quasi-fission process causes the wide charge distribution. The neutron-rich fragments such as $^{164,165}$Gd, $^{192}$W whose masses have not yet been measured experimentally, are observed in $^{154}$Sm+$^{160}$Gd at $b=1$ fm, and the production probabilities at $t=2000$ fm/c are about $2.5\times 10^{-3}$, $10^{-3}$ and $ 5\times 10^{-4}$ for $^{164}$Gd, $^{165}$Gd and $^{192}$W, respectively.

\begin{center}
\textbf{V. SUMMARY}
\end{center}

In this work, the improved quantum molecular dynamics (ImQMD) model is applied for the study of nearly-symmetric fusion reactions. By introducing the two-body inelastic scattering in the Fermi constraint procedure, the stability of an individual nucleus is further improved. The average numbers of spurious emitted nucleons are reduced by $50\%$ for $^{92}$Zr and $33\%$ for $^{132}$Sn comparing with those from the previous version of ImQMD, respectively. With this new version (v2.2), the fusion excitation functions of $^{16}$O+$^{186}$W and $^{132}$Sn+$^{48}$Ca have been re-examined. At energies around the fusion barrier (with $\sigma_{\rm fus}\simeq 10-100$ mb), the fusion cross sections are better reproduced with ImQMD-v2.2, because of the better initial nuclei being adopted. The fusion excitation functions of  $^{28}$Si+$^{30}$Si, $^{32}$S+$^{48}$Ca, $^{40}$Ca+$^{48}$Ca, and $^{86}$Kr+$^{76}$Ge can also be reasonably well reproduced with this semi-classical microscopic dynamics model. Simultaneously, we also investigate the smallest distance $R_{\rm min}$ between two nuclei in the back-angle quasi-elastic scattering events. The energy dependence of $R_{\rm min}$ is evidently observed, and the smallest surface separation between two nuclei from the back-angle quasi-elastic scattering events is about 2 fm at energies near the fusion barrier which is consistent with the value of the injection point distance, $s_{inj}$, adopted in the fusion-by diffusion model \cite{Cap14}.    

In addition, the quasi-fission process in $^{154}$Sm+$^{160}$Gd is also investigated with the microscopic dynamics model for the first time. We find that at energies above the Bass barrier, the fusion probability is smaller than $10^{-5}$, and almost all dinuclear systems tend to split up after $t=1500$ fm/c. For peripheral collisions, the quasi-elastic scattering plays a dominant role. Whereas for central collisions, the quasi-fission together with large charge and mass transfer plays a role to the wide charge distribution. From the central collisions of $^{154}$Sm+$^{160}$Gd, the neutron-rich fragments such as $^{164,165}$Gd, $^{192}$W whose masses have not yet been measured experimentally, can be produced according to the predictions of the ImQMD model. The quasi-fission reaction could be an alternative way to synthesize new neutron-rich heavy nuclei which is difficult to be produced with the traditional heavy-ion fusion reaction.

\begin{center}
\textbf{ACKNOWLEDGEMENTS}
\end{center}
This work was supported by National Natural Science Foundation of
China, Nos 11422548, 11275052, 11365005 and 11375062. N. W. acknowledges the support of the Open Project Program of State Key Laboratory of Theoretical Physics, Institute of Theoretical Physics, Chinese Academy of Sciences, China (No. Y4KF041CJ1).

\end{document}